\documentclass[onecolumn,preprintnumbers]{revtex4-2}
\usepackage{amssymb}
\usepackage{amssymb}
\usepackage{mathrsfs}
\usepackage{float}
\usepackage{bm}
\usepackage{amsmath}
\usepackage{mathdots}
\usepackage{graphicx}
\usepackage{subfigure}
\usepackage{array}
\usepackage{lipsum}
\usepackage{color}
\usepackage{hyperref}

\begin{document}
	
\title{On the Casimir effect from the zero-point energy: A tangential force and its properties}

\newcommand*{\SDUST}{Department of Physics and Institute for Theoretical Physics,\\Shandong University of Science and Technology, Shandong 266590, China}\affiliation{\SDUST}

\author{Zhentao Zhang}\email{zhangzt@sdust.edu.cn}\affiliation{\SDUST}

\begin{abstract}
We investigate the Casimir effect in the systems that consist of parallel but misaligned finite-size plates from the point of view of zero-point energy. We elaborate the zero-point energies of the radiation field in the perfect conductor systems would generate a tangential Casimir force, and explore the properties and consequences of this tangential force in various conductor systems. Thereafter, we generalize our discussion to dielectrics. After calculating the total zero-point energies of the surface modes in the multilayered systems, we show that the tangential force also exists in dielectrics. We obtain the finite-conductivity corrections to the tangential force for imperfectly conducting plates, and calculate the finite-temperature corrections to the force. The typical strength of the tangential force suggests it might be observable.
\end{abstract}

\maketitle

\section{Introduction}

In quantum field theory, the equal-time commutation relations of the radiation field operators would result in divergent constant in the free Hamiltonian, and the divergent constant is interpreted as the zero-point energy of the quantum field or energy of the vacuum; it affects nothing but gravity, see, e.g.,~\cite{Weinberg}. Nevertheless, the change of the vacuum energy can be measurable if we disturb the vacuum by imposing special boundary conditions on the quantum field. This is the established understanding for the Casimir effect \cite{Casimir} in particle physics, see, for example,~\cite{Weinberg,Itzykson,Zee}. Though there is a controversy about the origin of the Casimir effect in the literature~\cite{Jaffe}, we shall consider the Casimir effect mainly from viewpoint of the zero-point energy.

To explore the Casimir effect in the context of zero-point energy, we shall discuss the physical systems in which the Casimir forces predicted by the zreo-point energies may have special properties. The systems considered by us will be paralleled perfectly conducting plates, but the plates are finite-size and will be misaligned. In these systems the zero-point energies of the radiation field would require that, besides the normal Casimir forces, the plates should experience the tangential forces. Compared with the normal Casimir force, the tangential force may have interesting properties. Thereafter, we shall generalize our study to dielectrics.

The Casimir physics is an active multidisciplinary research area and in recent decades theoretical and experimental progresses has been made in different aspects of this field, see, for review, \cite{Plunien,Milonni,Kardar,Milton,Bordag1,Klimchitskaya,Bordag2,Dalvit,Woods}. To our knowledge, focusing on the edge correction to the Casimir effect, a similar situation has been studied in the multiple scattering formalism~\cite{Wagner,Milton2} for a massless scalar field, which does not involve the concept of the zero-point energy of the scalar field. Thus, the present paper should not be considered as a work that generally remarks tangential interactions may occur in the system that consists of flat plates. Instead, one part of our study may be considered as a specific investigation on the various misaligned conductor systems from the point of view of the zero-point energy for the radiation field; the approach would produce results in simple forms and in the meantime we shall show that they are sufficient to provide valuable information on the systems, and the other part would be a generalization for dielectric systems.

\begin{figure}[H]
	\centering
	\includegraphics[width=8.5cm]{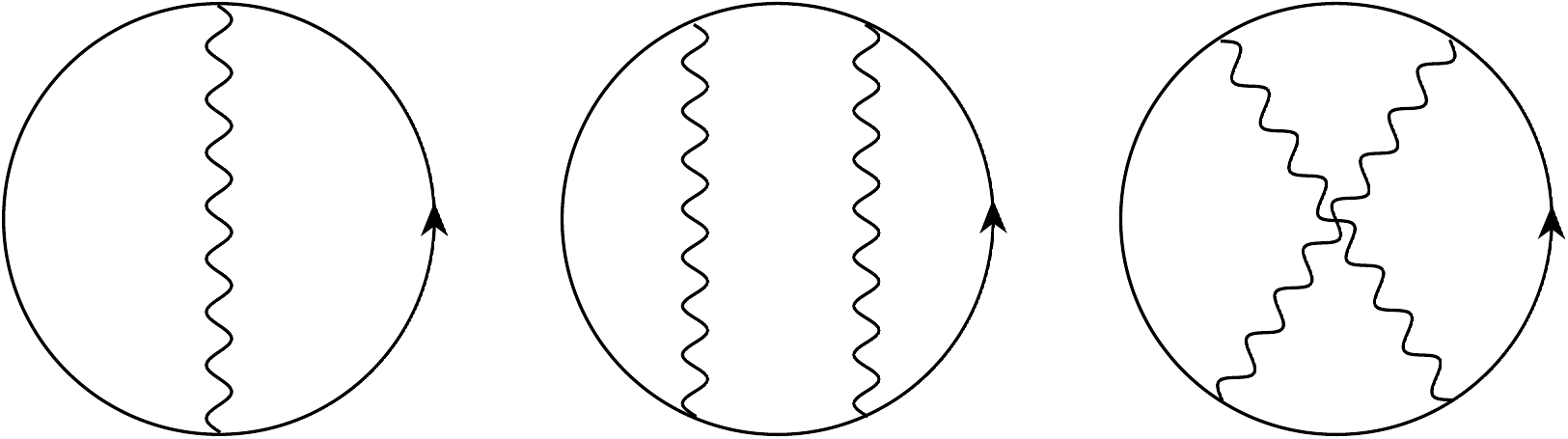}
	\caption{Some representative disconnected diagrams for the quantum electrodynamics vacuum fluctuations in perturbation theory.}
	\label{Bubble}
\end{figure}

Due to the multidisciplinary feature of the Casimir physics, occasionally, the physical meaning behind a terminology can be subtle for readers with different backgrounds. Thus, we would like to eliminate potential ambiguity at the beginning: We would stick to the standard description of the Casimir effect in quantum field theory. This means we shall discuss the free Maxwell theory in confined configurations~\cite{Weinberg,Itzykson,Zee}. We may also mention that when we apply perturbation theory to the interacting quantum fields, there is the other type of divergent vacuum energy emerging at the loop-level: the ``vacuum bubbles'', see Fig.~\ref{Bubble}. However, it is well known that these disconnected vacuum diagrams contribute to neither the Casimir effect nor the S-matrix, see, e.g.,~\cite{Weinberg, Itzykson,Zee}. Thus, we shall restrict ourselves to the zero-point energy of the radiation field and discard the energy from the virtual particles fluctuations in the vacuum.

\section{The tangential forces} 
\begin{figure}[H]
	\centering
	\includegraphics[width=8.5cm]{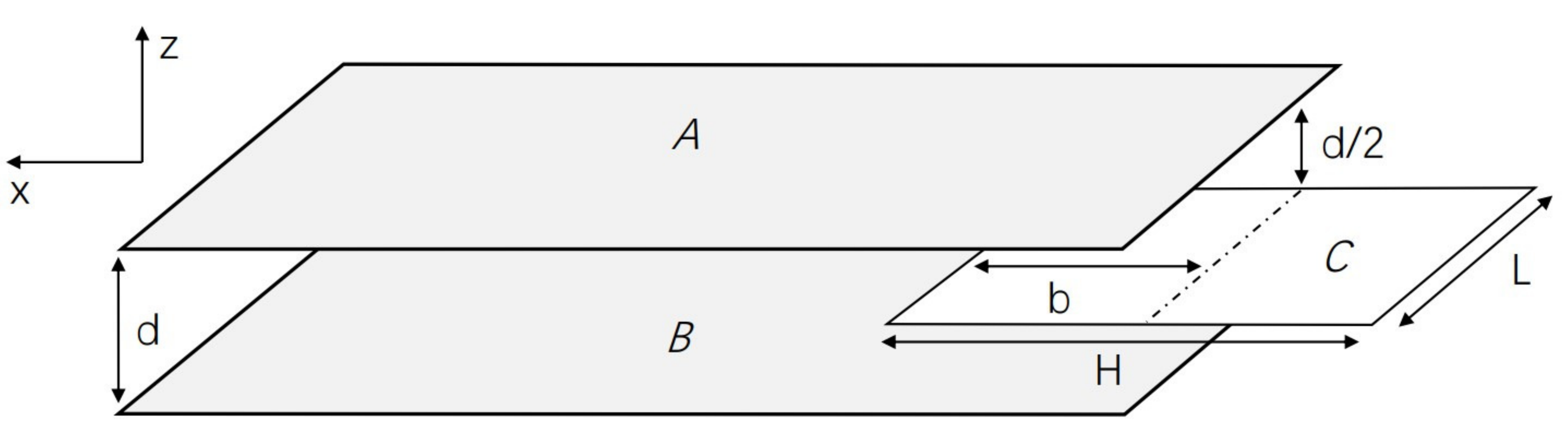}
	\caption{The conducting plate C lies partly in the middle of two parallel plates. The small distance between the two larger plates is $d$.}
	\label{Fig2}
\end{figure}

Consider a physical system that consists of three macroscopic~\cite{Note1} parallel rectangular conducting plates, see Fig.~\ref{Fig2}. There is a thin plate C lying partly in the middle of two identical plates A and B, and to facilitate the discussion, we shall assume its dimension is smaller than plate A; its width is $L$ and length is $H$. The overlapping area of plate C and the adjacent plates is $Lb$, where $L, b, H-b\gg d$.

\begin{figure}[H]
	\centering
	\includegraphics[width=8.5cm]{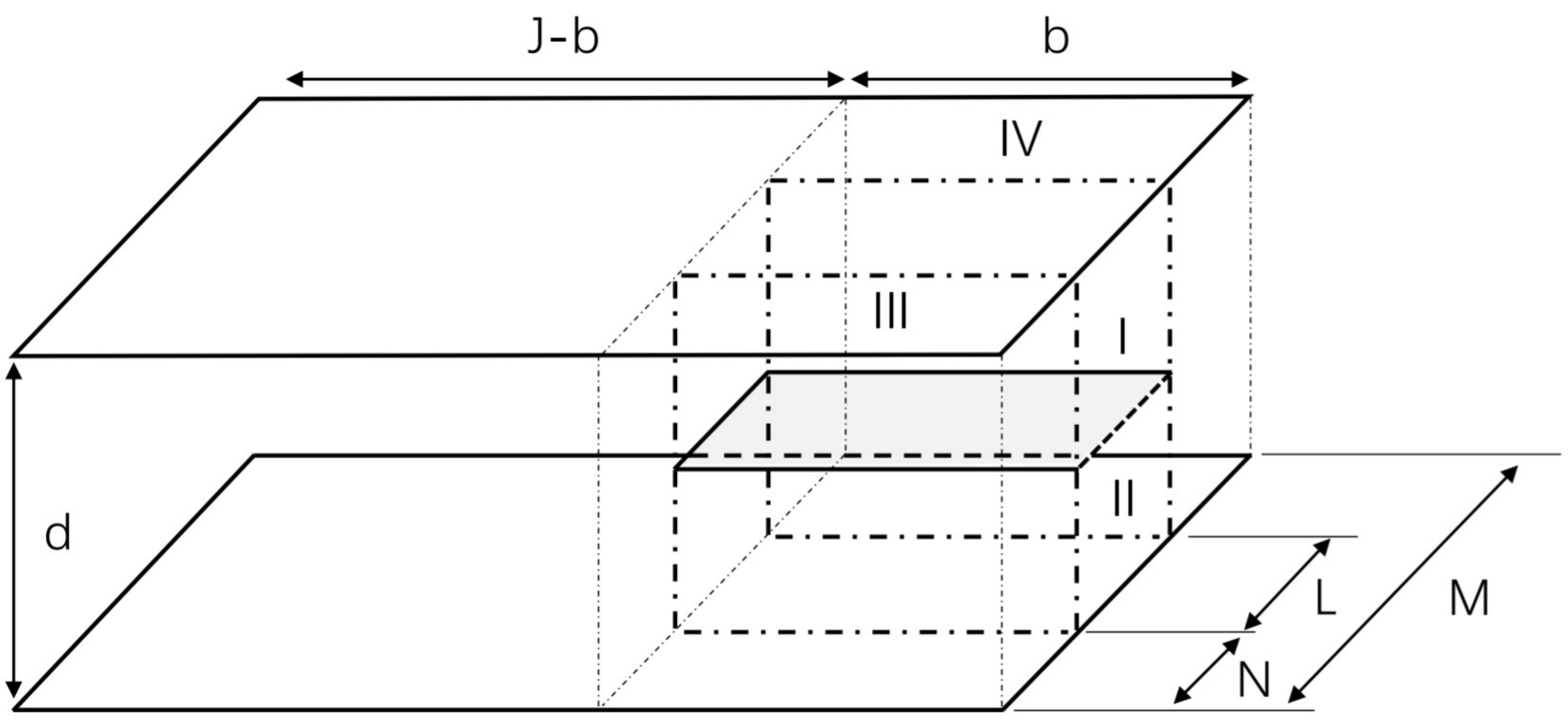}
	\caption{We divide the configuration of the system into the sub-configurations, where the outside part of plate C is ignored, and $N,M-N-L,J-b \gg d$.}
	\label{Fig3} 
\end{figure}

For this system, we can understand that the zero-point energy would predict the normal Casimir force experienced by plate C is zero. However, let us consider the system more carefully and calculate explicitly the zero-point energy of this system. Usually, a small modification of a simple configuration would increase considerably the difficulty in calculating the zero-point energy of the radiation field. Though various theoretical methods were introduced to Casimir physics for calculating the Casimir force, we shall consider the Casimir effect from the viewpoint of zero-point energy and we may study this nontrivial configuration in the simple way: Dividing the configuration into sub-configurations, see Fig.~\ref{Fig3}.  

First, we should justify here this intuitive division in case it is not well-defined. We know that the bare quantity of the zero-point energy of the radiation field for the configuration $I+II$ is
\begin{align}
&E^{I+II}_{0}=\frac{\hbar c}{2}\int Lb\frac{d^2k_{\Arrowvert}}{(2\pi)^2}\left[k_\Arrowvert+2\sum_{n=1}^{\infty}\left(k_\Arrowvert^2+\frac{n^2\pi^2}{d^2}\right)^{\frac{1}{2}}\right],
\end{align}
where the symbol $I+II$ represents the configuration integrated by the sub-configurations $I$ and $II$ when plate C does not exist, and for the macroscopic scale parameters $N$ and $M-N-L$, we have 
\begin{equation}
E^{III}_{0}=E^{I+II}_{0}(L\longrightarrow N)
\end{equation}
and
\begin{equation}
E^{IV}_{0}=E^{I+II}_{0}(L\longrightarrow M-N-L),
\end{equation}
where $L\rightarrow N$ and $L\rightarrow M-N-L$ indicate variable $L$ is changed to the latter variables, $k$ is the wavenumber, and $k_\Arrowvert^2+n^2\pi^2/d^2=k^2$. Then, the whole zero-point energy in the sub-configurations is
\begin{align}
&E^{I+II}_0+E^{III}_0+E^{IV}_0\nonumber\\
&=\frac{\hbar c}{2}\int Mb\frac{d^2k_{\Arrowvert}}{(2\pi)^2}\left[k_\Arrowvert+2\sum_{n=1}^{\infty}\left(k_\Arrowvert^2+\frac{n^2\pi^2}{d^2}\right)^{\frac{1}{2}}\right]\nonumber\\
  &=E^{I+II+III+IV}_{0},
\end{align}
where $E^{I+II+III+IV}_0$ represents the bare zero-point energy of the corresponding integrated configuration when plate C does not appear. One can see that our division of the configuration will not change the vacuum energy of the system. Thus, the intuitive division may be well-defined, and this approximation method for finding the zero-point energy would be qualified to provide valuable information on the system.

Now let us consider the zero-point energies of the sub-configurations $I$ and $II$ separately. Clearly,
\begin{align}
	E^{I}_{0}=E^{II}_{0}=\frac{\hbar c}{2}\int Lb\frac{d^2k_{\Arrowvert}}{(2\pi)^2}\left[k_\Arrowvert+2\sum_{n=1}^{\infty}\left(k_\Arrowvert^2+\frac{n^2\pi^2}{(d/2)^2}\right)^{\frac{1}{2}}\right],
\end{align}
where $k_\Arrowvert^2+n^2\pi^2/(d/2)^2=k^2$.

Comparing with the zero-point energy of the configuration $I+II$ when plate C does not exist, the vacuum energy difference is 
\begin{equation}
E=E^{I}_{0}+E^{II}_{0}-E^{I+II}_{0}.
\label{diference1}
\end{equation}
To find the difference from the divergent quantities, numerous regularization methods can be employed, see, for example, \cite{Itzykson,Plunien}. We may find that the finite energy difference is
\begin{align}
	E=-\frac{\pi^2\hbar c}{48d^3}Lb.
	 \label{Zero-point1}
\end{align}

Notice that if plate C moves along the $x$-direction, the motion will alter the original configuration and the vacuum energy of the system will also be changed. Therefore, it turns out that the conducting plate in the system may experience a net Casimir force $F$ along the $x$-direction
\begin{equation}
	F=F_T=-\frac{\partial E}{\partial b}=\frac{\pi^2\hbar c}{48d^3}L.
	\label{tangential}
\end{equation}
This Casimir force depends on the length $L$, not the overlapping area which would generally be a variable for the Casimir forces of the other physical systems. It is an attractive force which will drag the plate into the space between plates A and B. For the ideal boundary conditions of the sub-configurations, the force disappears when the plate fully goes into the space, and the plate would maintain a constant velocity to reach the other side~\cite{Pendry}. Thereafter, an opposite Casimir force may be turned on, and the velocity of the plate goes to zero and then the plate will turn around. Thus plate C oscillates in the plane.

We may also notice that unlike the normal Casimir force, due to the $1/d^3$-dependence on the small separation, the strength of this Casimir force is weak in general. However, from a modern mechanics point of view, the force might be large enough to be detected. For example, for a moderate setting $d=0.3~\mu\text{m}$ and $L=0.02$~m, the strength of the Casimir force experienced by plate C is
\begin{equation}
	F=4.8\times10^{-9}~\text{N}.
\end{equation}

\begin{figure}[H]
	\centering
	\subfigure[The vertical view for plate C. For $L\sim10^{-2}$~m, $L_{in}=L$ effectively, due to the fact that the typical distance $d\sim10^{-7}$~m.]{\label{regular}\includegraphics[width=5.6cm]{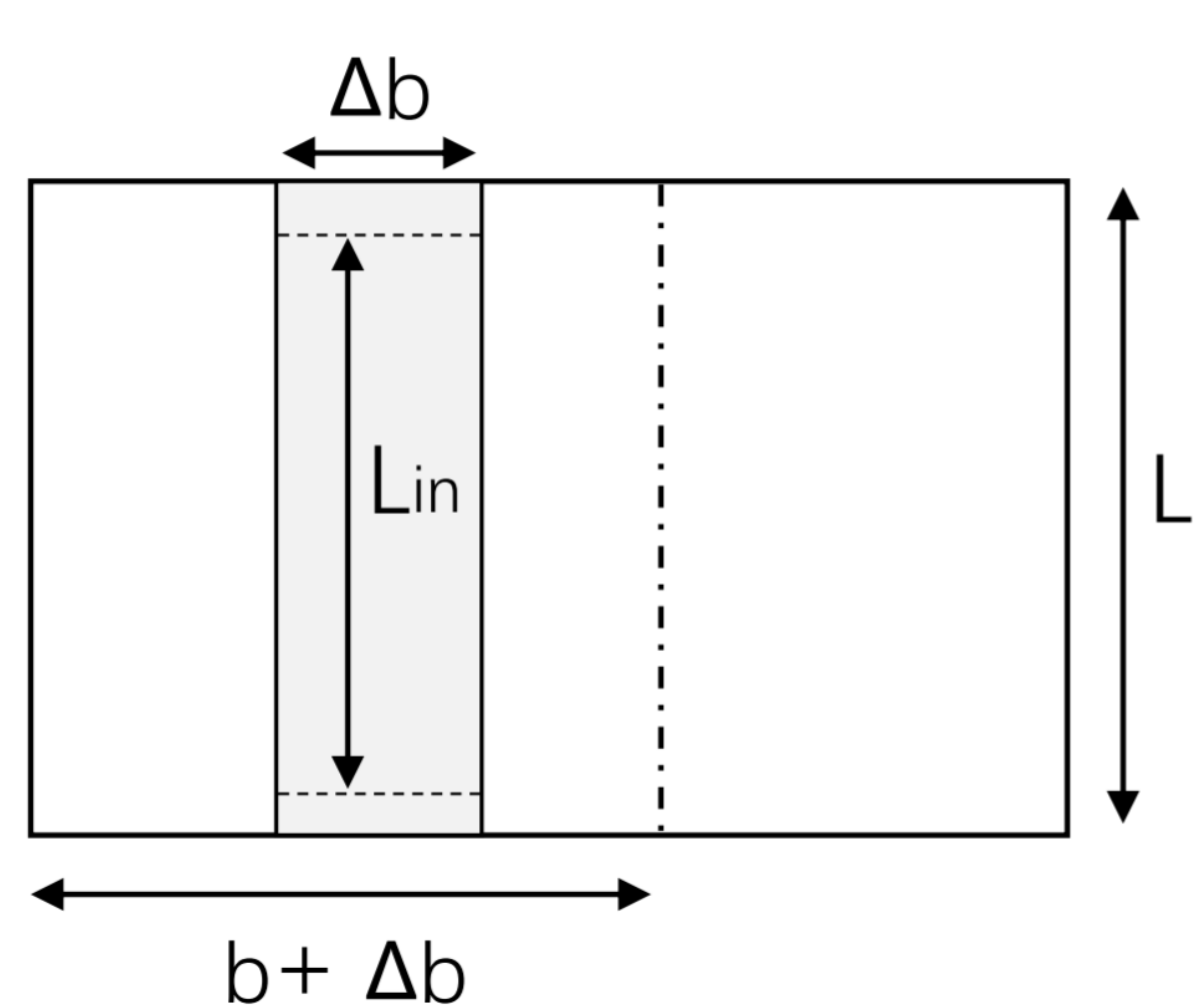}}
	~~~\subfigure[A conducting plate with an irregular edge.]{\label{irregular}\includegraphics[width=5.9cm]{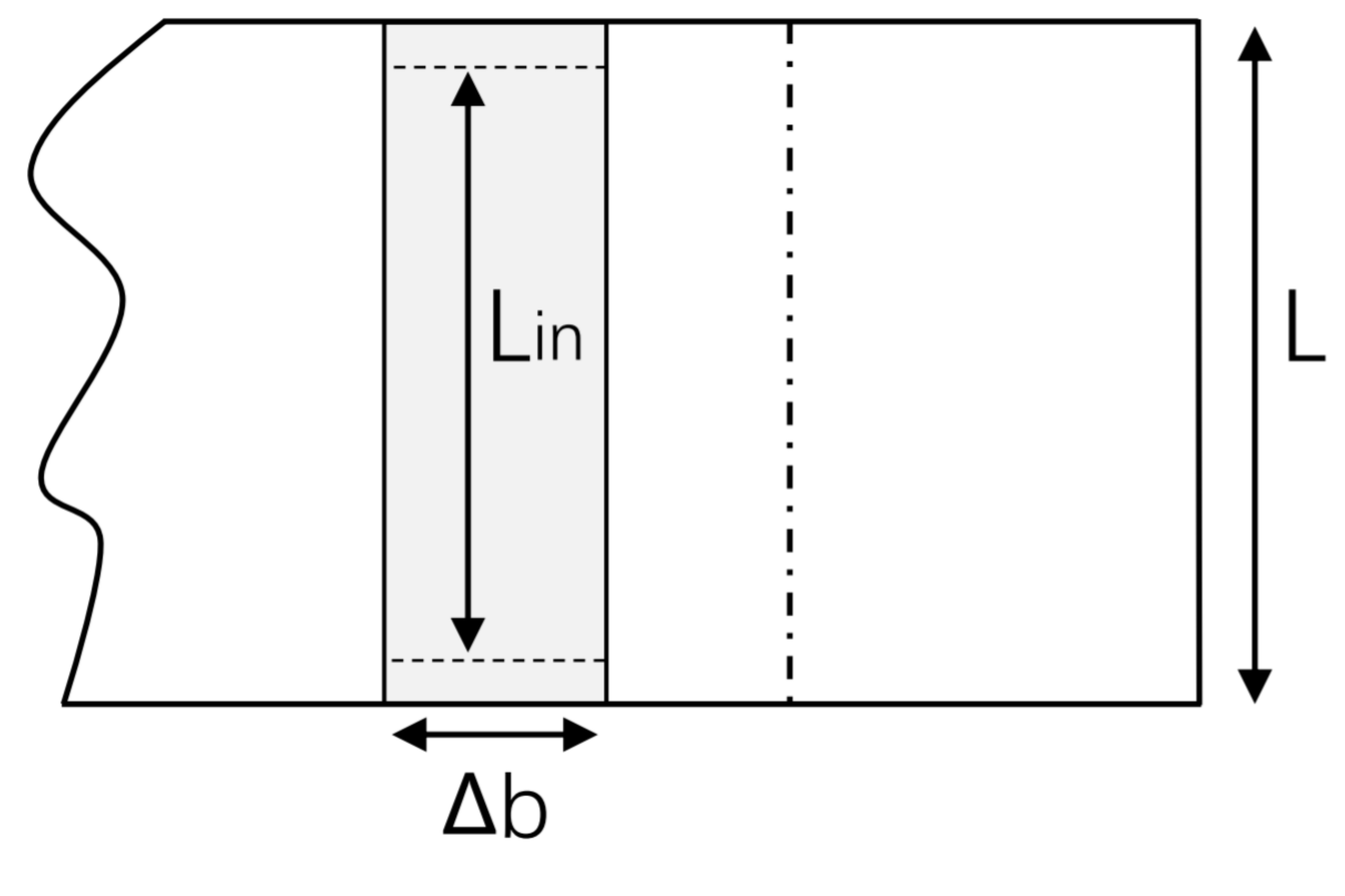}}
    \caption{The vertical view for the conducting plates.}
\end{figure}

Before we further discuss the implications of this tangential Casimir force, however, a remark should be given here. One may notice that to obtain the major physical properties of the system, in the above discussion we have used the ideal boundary conditions for the parallel plates. This means there would be a dramatic change of the sub-configurations for the vector field near the edges, i.e., two adjacent space points near an edge may belong to different sub-configurations. While it can be an ideal model in our field-theoretical approach, can there be any hidden dangers in this assumption, which may greatly weaken our conclusion? The answer is negative. We may emphasize that for this Casimir force, we focus on the change of the overlapping area, which would induce the change of the vacuum energy. An infinitesimal change in the $x$-direction $\Delta b$ will not affect the physical boundary situation for the frontal edge as long as $J-b$ maintains a macroscopic scale, and then we can consider the change happens in the inner region of the overlapping area~\cite{Note2}. And the potential corrections from the additional side edges may be ignored, since the dominant contribution to the change of the vacuum energy comes from the interior zone of $\Delta b L$, see Fig.~\ref{regular}. Similarly, this analysis can also be applied to a plate with imperfect edges or the geometry shown in Fig.~\ref{irregular}. An estimation for the precision of Eq.~(\ref{tangential}) affected by the geometry may be given here. Since the transition region for the energy density near an edge $\lesssim d/2$~\cite{Note2}, the precision of the analytic formula may reach $\sim4\times d/(2L)=2d/L$. For $L\sim10^{-2}$~m and $d\sim10^{-7}$~m, the precision $\sim10^{-5}$. (It may be noted that when the distances between the plates are $d/2$, $R-d/2$, and $R$, the condition $L\gg \text{min}\{d/2,R-d/2\}$ would be sufficient for our ideal boundary approximation.)

However, it might be useful to note that if one wants to apply the analytic result to the system where $\arrowvert b\arrowvert\lesssim d$ or $\arrowvert H-b\arrowvert\lesssim d$, a non-negligible positive correction from the physical boundaries to the ideal-boundary formula should be expected. In these two small geometric parameter regions, numeric methods should be employed. We here content ourselves with the analytic expression, since for most accessible geometric parameter regions in experiments, the analytic formula will provide reliable information on the system.

We should also note here that it is easy to notice the properties of this Casimir force experienced by the flat plate may be different to the lateral Casimir force caused by the corrugated surfaces of the plates~\cite{Golestanian1,Golestanian2,Chen1,Chen2}.

\begin{figure}[H]
	\centering
	\includegraphics[width=8cm]{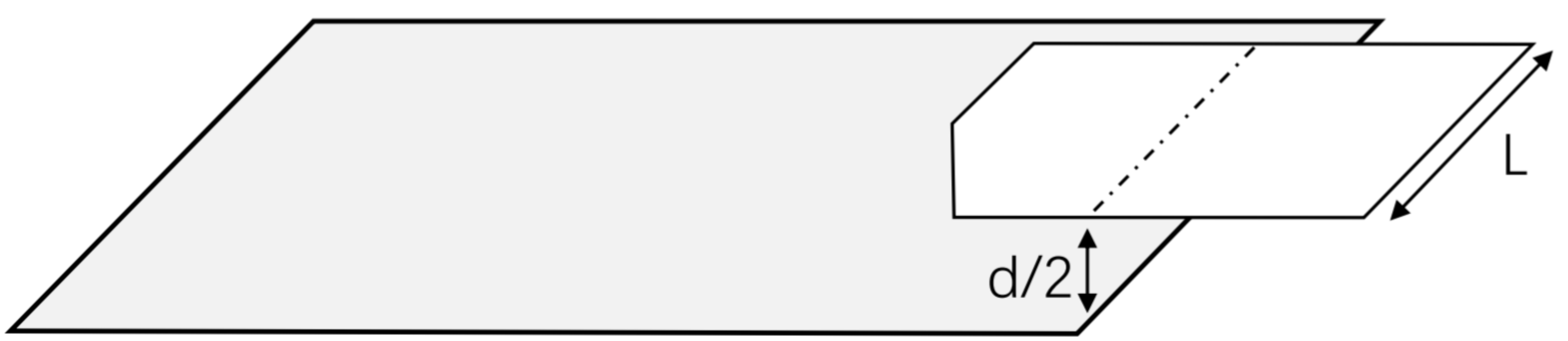}
	\caption{The system consists of two parallel conducting plates separated by $d/2$. The upper plate may have an irregular edge.}
	\label{Fig5} 
\end{figure}

We have shown that the strength of the tangential Casimir force might be measurable, but one may worry about the feasibility of this type of experiment, since the typical distance between two plates in Casimir physics is sub-$\mu$m and it might be difficult to build a thin conducting plate which can be put in the middle of the two plates. However, to measure the tangential force, a system that consists of two finite-size conducting plates may also be used (see also \cite{Wagner}), see Fig.~\ref{Fig5}. This system can be considered as the situation that we move the conducting plate A or B in the previous system to infinity, and following the discussion given above, we may find easily the tangential force between the plates
\begin{equation}
	F'_T=\frac{\pi^2\hbar c}{90d^3}L.
	\label{2}
\end{equation}
                 
But notice that in this system the normal Casimir force $F'_N$ also exists, and for $b=0.01$~m and $ d=600$~nm, $ F'_N=10^{5} F'_T$ for the case of two rectangular plates. Due to the appearance of this large normal force, one can understand that measuring the tangential force in this system may still need cutting-edge experimental technologies. Nevertheless, the situation that the directions of them are perpendicular may be in our favor, since we do not need to use the large normal force as a baseline for extracting the information of the tangential force in this kind of measurement. Notice also that the so-called tangential force here is the tangential component of the Casimir force, and the Casimir force in this system is not independent of the overlapping area between the plates.

We find here the interesting relation
\begin{equation}
F'_{T} > \frac{1}{2} F_{T},
\label{neq}
\end{equation}
which indicates that conducting plate C feels the tangential force caused by one plate is not half the force caused by two plates when the two plates maintain an exact spatial symmetry for plate C. From the general theory of van der Waals force, we know that the Casimir force is not additive~\cite{Dzyaloshinskii}, which means the macroscopic Casimir force may not be found by adding up all the microscopic van der Waals forces for the atoms or molecules in the plates. Similarly, Eq.~(\ref{neq}) shows the tangential Casimir force is not additive even for the macroscopic objects. From the perspective of quantum field theory, the explanation for this inequality can be simple: It is illegal to consider that the tangential force $F_T$ consists of two independent tangential forces caused separately by plates B and C; the force is inseparable, since it is induced by the configuration of the system in a unified way.

\section{The tangential forces in other conductor systems}
The tangential force from the radiation field, in general, exists in the systems that consist of misaligned finite-size plates, and we would like to further discuss the tangential Casimir force in other systems which might have instructive properties. 

\begin{figure}[H]
	\centering
	\includegraphics[width=8cm]{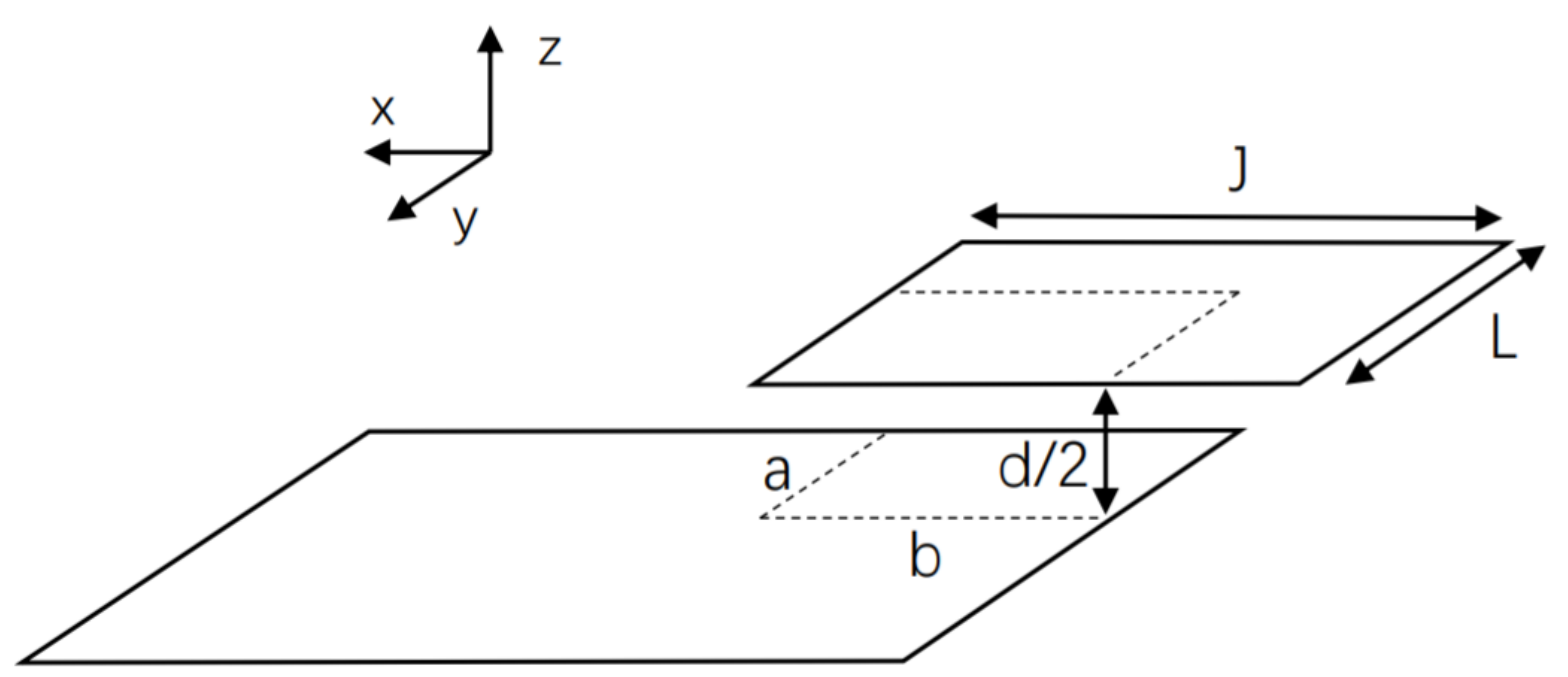}
	\caption{The system consists of two parallel conducting plates separated by $d/2$, where the overlapping area between them is $a\cdot b$, and $a, b, L-a, J-b\gg d$.}
	\label{Fig7} 
\end{figure}

We may consider an obviously misaligned conducting system shown in Fig.~\ref{Fig7}.
In this system both of $a$ and $b$ are the variables for the zero-point energy of the radiation field. Thus, the tangential force would have two nonzero components in Cartesian coordinates, and the tangential force experienced by the upper plate is
\begin{align}
	{\textbf{\textit{F}}}_T=\frac{\pi^2\hbar c}{90d^3}\left( a{\bf{e}}_{x}+b{\bf{e}}_{y} \right).	
\end{align}
Notice that both of the direction and strength of this force will change if the plate can be driven by the tangential force. (If we replace the rectangular plates with other parallelogram plates, the tangential force may also be easily found.) In this place we might note that from the same arguments given in the last section, we may understand that when the smaller plate is just draged into the upper space of the larger plate, the tangential force will not disappear immediately. But the force would decrease rapidly if the tangential distances between the ``external'' edges of the smaller plate and edges of the larger plate are increased. For the tangential distances which are bigger than $d/2$, one may consider the tangential force is effectively zero~\cite{Note}.

\begin{figure}[H]
	\centering
	\includegraphics[width=8.5cm]{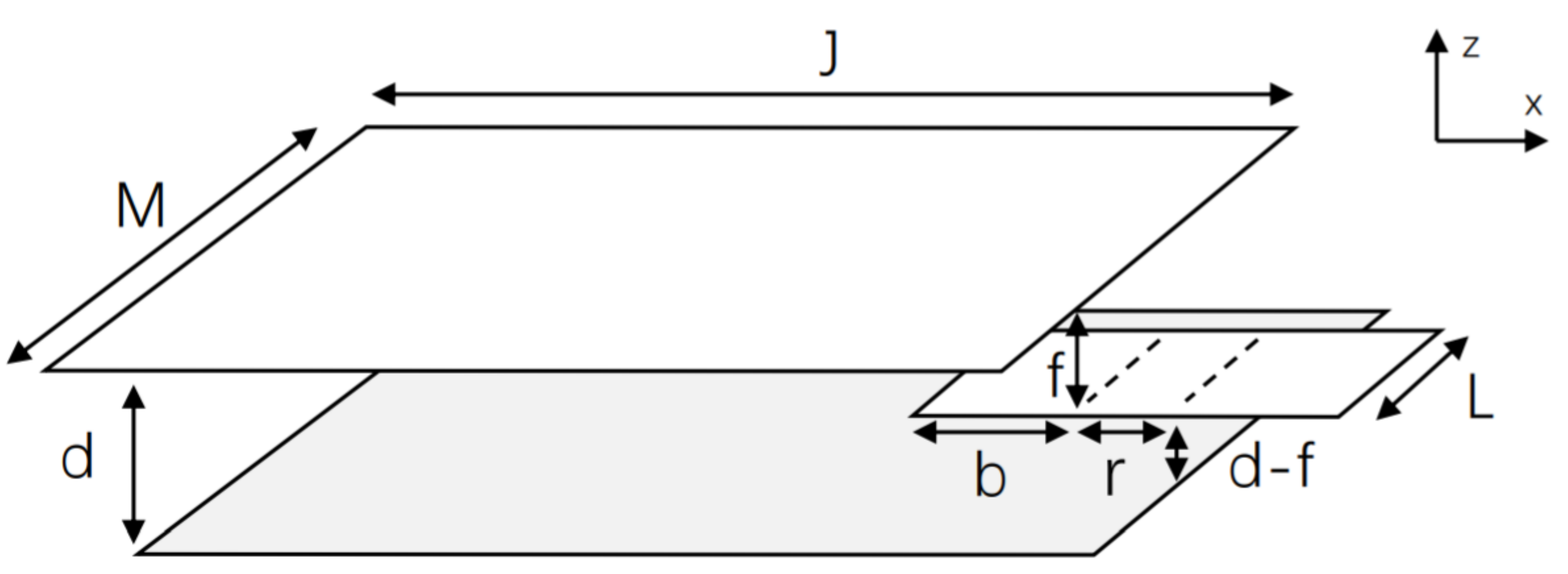}
	\caption{The system consists of three parallel conducting plates. The geometries of the upper and bottom rectangular plates are the same but they are misaligned in the $x$-direction. All $d$-irrelevant parameters in the system are greatly larger than the small distance $d$.}
	\label{Fig8} 
\end{figure}

Now let us consider a more complicated but interesting system, see Fig.~\ref{Fig8}, and without loss of generality, one may assume the distances between the conducting plates are fixed.
To calculate the zero-point energy of the radiation field in this system, we may also divide the configuration into sub-configurations and various regularization methods could be employed. We may find the vacuum energy difference in this system is
\begin{equation}
E=-\frac{\pi^2\hbar c}{720}\left[\frac{J-r}{d^3}M-\frac{b}{d^3}L+\frac{b}{f^3}L+\frac{b+r}{(d-f)^3}L\right].
\end{equation}
Then the tangential force felt by the bottom plate is 
\begin{equation}
F^{bottom}_T=-\frac{\partial E}{\partial r}=\frac{\pi^2\hbar c}{720}\left[\frac{L}{(d-f)^3}-\frac{M}{d^3} \right].
\end{equation}
We can see that the direction of the force is changeable, and the equilibrium condition for the parameters is $L{d^3}=M(d-f)^3$.

As we should expect, one may notice that the sum of the tangential forces of the three plates is zero, since the tangential force experienced by the upper plate is
\begin{equation}
	F^{upper}_T=\frac{\pi^2\hbar c}{720}\left[\frac{L}{f^3}+\frac{M-L}{d^3} \right],
\end{equation}
and the middle plate experiences the tangential force
\begin{equation}
	F^{middle}_T=\frac{\pi^2\hbar c}{720}L\left[\frac{1}{d^3}-\frac{1}{f^3}-\frac{1}{(d-f)^3} \right].
\end{equation}

\begin{figure}[ h]
	\centering
	\subfigure[Two circular plates separated by a small distance $d$.]{\label{rounda}\includegraphics[width=5.5cm]{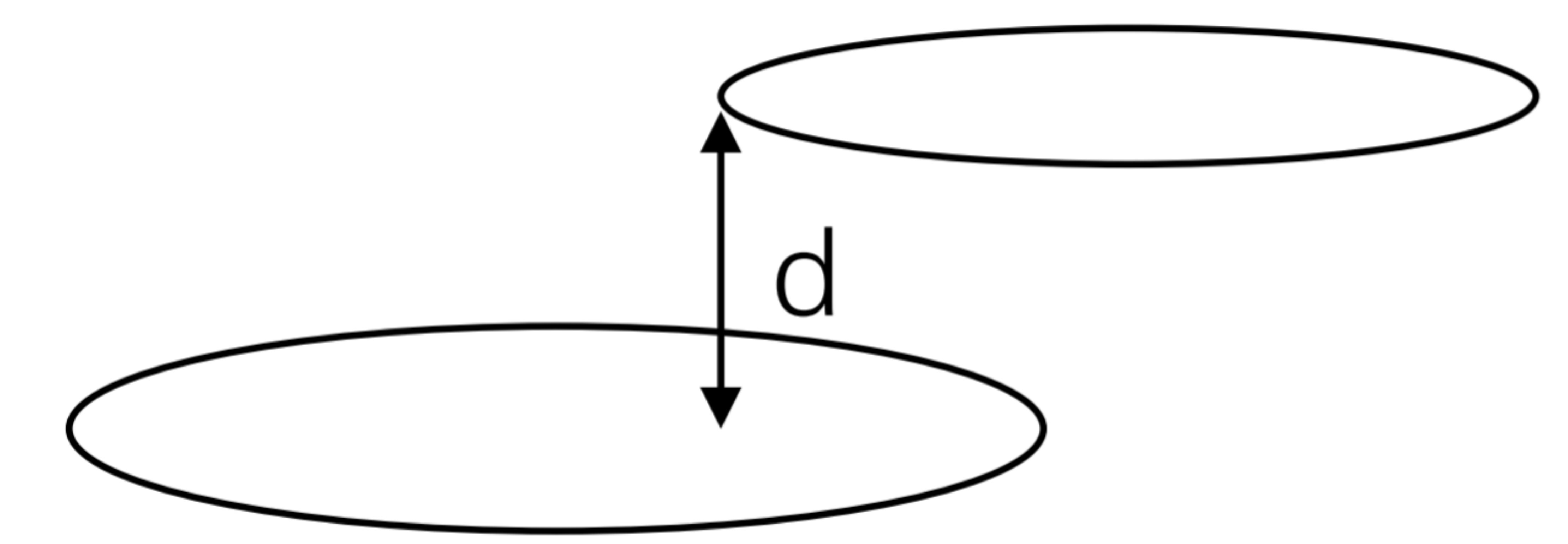}}
	~~~\subfigure[The vertical view for the plates. The radii of the plates are $R_1$ and $R_2$. The distance between their centres is $b$ and $R_1-R_2<b<R_1+R_2$.]{\label{roundb}\includegraphics[width=5.5cm]{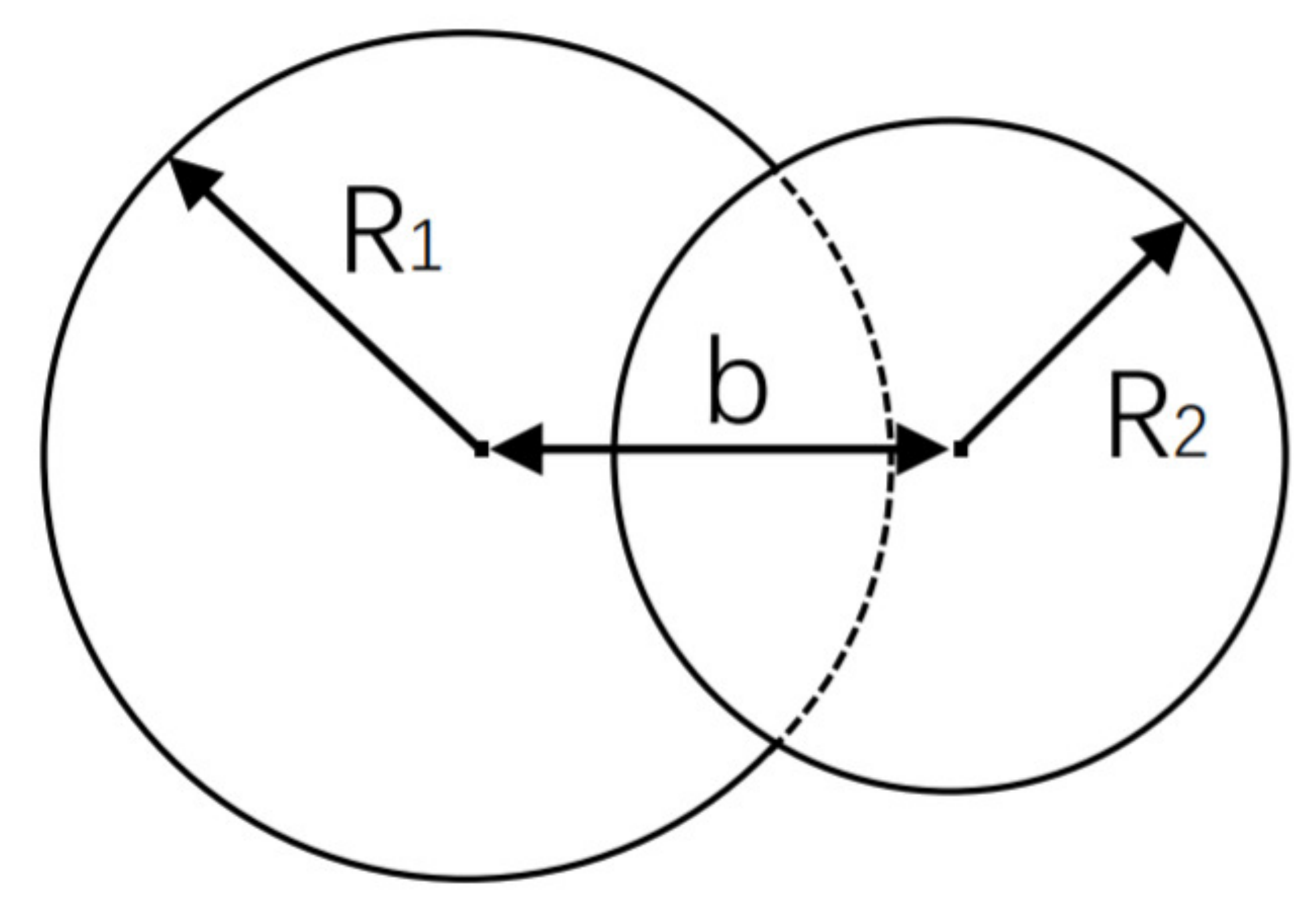} }
	\caption{The misaligned system consists of two parallel conducting circular plates, where $d\sim 10^{-7}$~m and $R_{1,2}\sim 10^{-2}$~m.}
	\label{round}
\end{figure}

Circular plate is also an interesting object in Casimir physics, and it may be of interest to consider the misaligned systems that consist of a rectangular plate and a circular plate or two circular plates. We consider here a representative system in this context, see Fig.~\ref{round}. 

\begin{figure}[H]
	\centering
	\includegraphics[width=8.5cm]{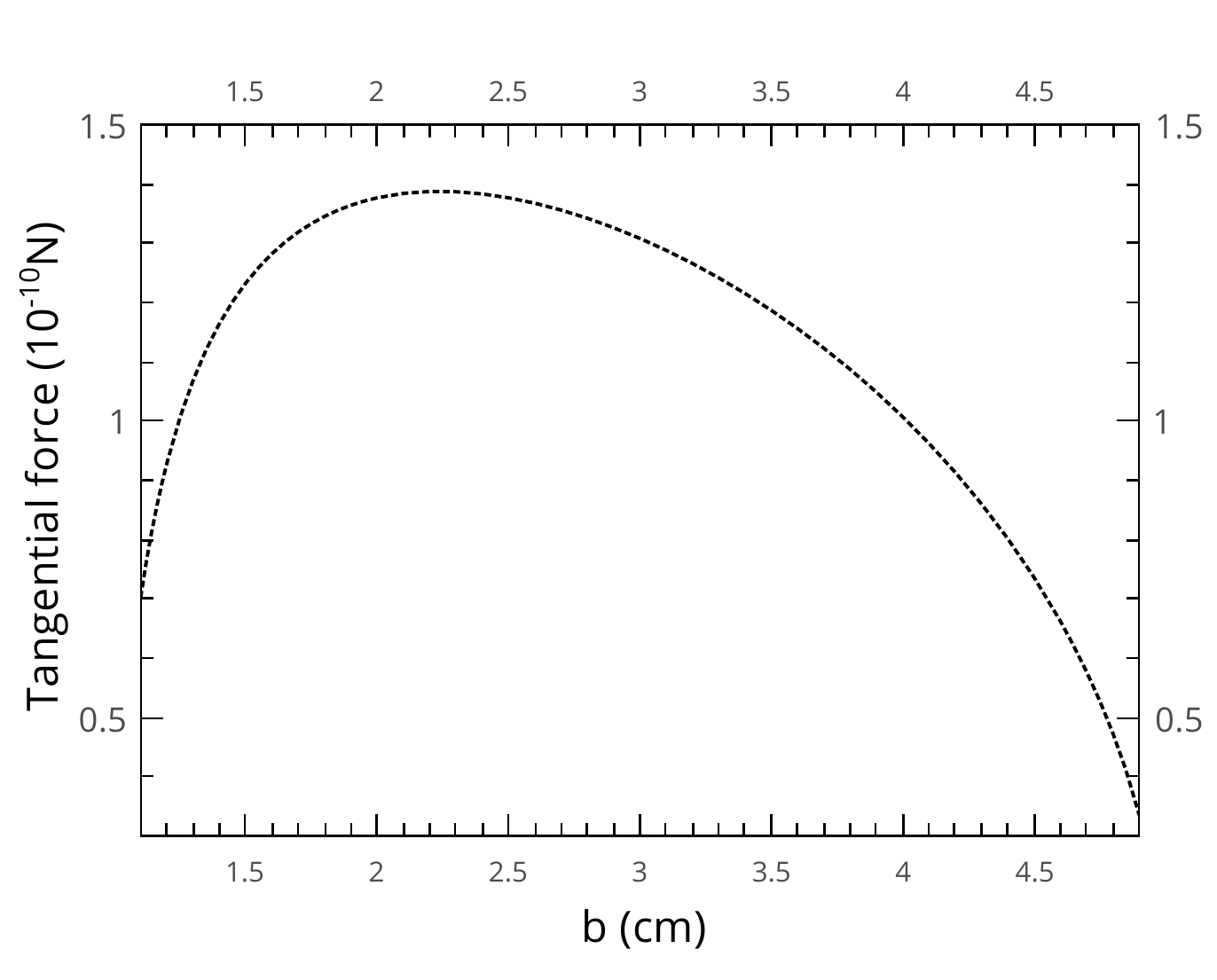}
	\caption{The tangential force between the circular plates, where the distance $d=500$~nm.}
	\label{R}
\end{figure}

For this conductor system, we may find that the tangential force generated by the zero-point energy of the radiation field is 
\begin{equation}
	F_T=\frac{\pi^2\hbar c}{720bd^3}\sqrt{4b^2R^2_1-(R^2_1-R^2_2+b^2)^2},
\end{equation}
and the directions of the attractive tangential forces experienced by the two circular plates are along the straight line connecting their centres. For $R_1=3$~cm and $R_2=2$~cm, the tangential force is plotted as a function of the distance between the centres, see Fig.~\ref*{R}.

\section{The tangential forces in dielectrics}

From the viewpoint of quantum field theory, one might be inclined to consider that the Casimir effect may disappear if the conducting plates are replaced with dielectric plates, since the plates may no longer impose an explicit boundary condition on the quantum field. However, the Casimir force does exist in dielectrics~\cite{Lifshitz}, and in this context the Casimir force is generally understood as the macroscopic nontrivial realization of the van der Waals forces between atoms or molecules in the plates. But one can also discuss the Casimir effect in dielectrics from the viewpoint of zero-point energy for the surface modes~\cite{Milonni,van Kampen}.

In this section we generalize our discussion to dielectrics. The basic idea may be the same: We shall calculate the zero-point energies in the misaligned systems with the ideal geometric boundary conditions, which will be a reliable approximation for the situation considered by us, and show the energy difference would induce a tangential force between the dielectric plates. Our discussion will follow closely the zero-point energy formalism developed in~\cite{Milonni}.

\subsection{A zero-point energy approach}

\begin{figure}[H]
	\centering
	\includegraphics[width=8.5cm]{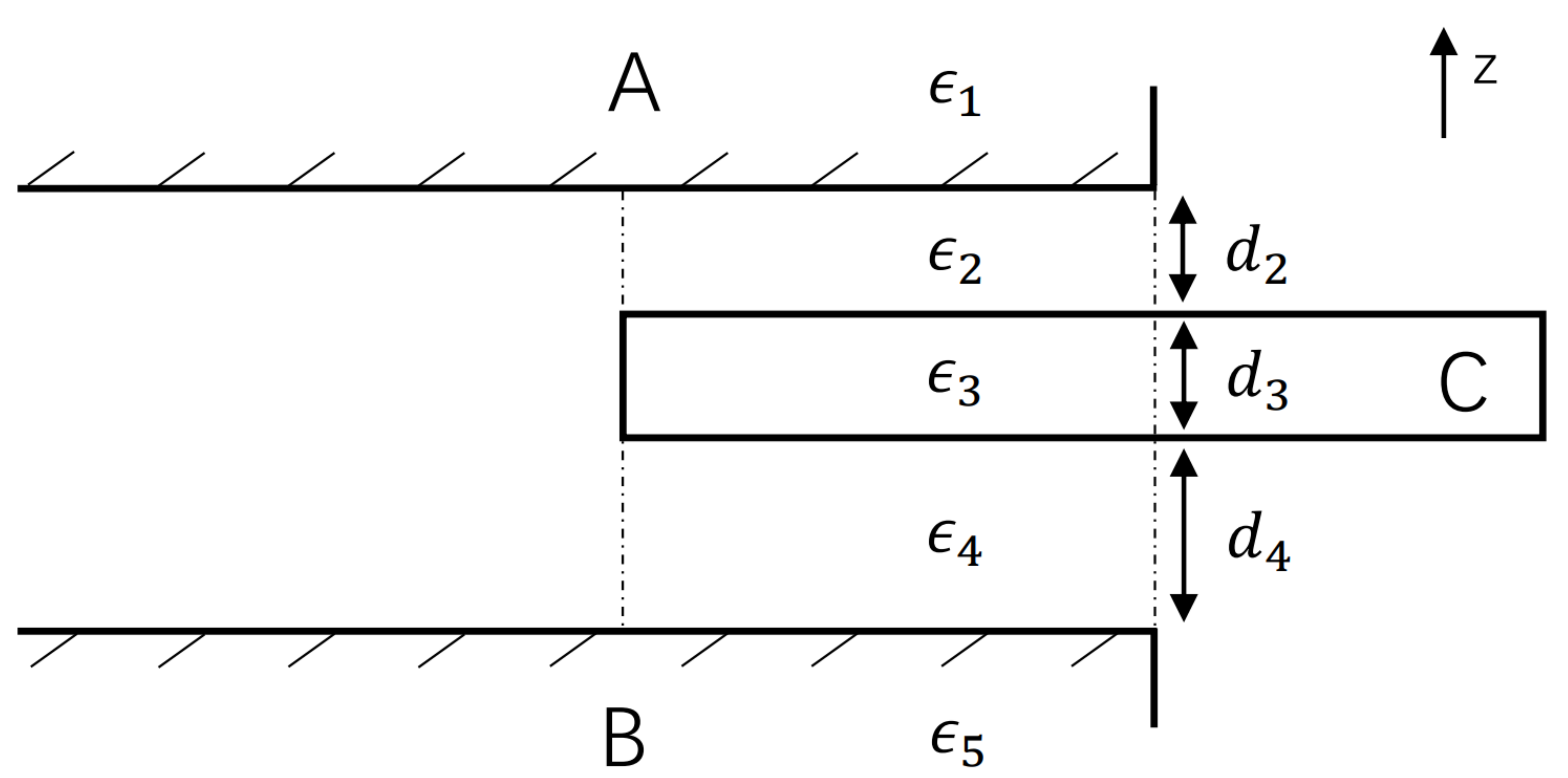}
	\caption{The side view for the misaligned dielectric systems, where a smaller plate is partially inserted into the space between two thick plates. All the horizontal parameters for this system are the same as in Fig.~\ref{Fig2}(\ref{Fig3}).}
	\label{slab}
\end{figure}

We now consider a variant system of Fig.~\ref{Fig2}(\ref{Fig3}), see Fig.~\ref{slab}. We assume $L\gg \text{min}\{d_2,d_3,d_4\} (\sim 10^{-7}~\text{m})$ and that $b,H-b$ are significantly larger than $\text{min}\{d_2,d_3,d_4\} $. In general, these geometric conditions would be met naturally by experiments in this type unless one designs an experiment to investigate the boundary effect. From the arguments and the numerical simulations cited in~\cite{Note2}, one may understand that even the infinitely thick plate would not effectively alter the precision of ideal boundary conditions estimated before for a thin plate (For example, for the perfect conductors, the situation of an infinitely thick plate would give around $23\%$ correction to the edge-correction contribution from a thin plate~\cite{Note2}).

For the misaligned system, we are interested in the difference of the zero-point energies caused by the appearance and absence of plate C between slabs A and B. For this purpose, we need to calculate the zero-point energy of the surface modes in five macroscopic-scale dielectric layers. Implications of the boundary conditions for the surface modes in the five regions has been explored by~\cite{Zhou}. However, the zero-point energy obtained in~\cite{Zhou}, see also~\cite{Klimchitskaya2000}, is not suitable for the present study, since the energy of the system would go to zero if the thickness of the middle plate approaches infinity ($d_3\rightarrow\infty$) and then it may not represent the total energy of the multilayered system. 

The boundary conditions for the electromagnetic field required at (say) $z=0,d_4,d_3+d_4,d_2+d_3+d_4$ in the five layers~\cite{Zhou} would generate two systems of eight homogeneous linear equations, and it may be found that the existences of non-trivial solutions of these equations require
\begin{align}
	G^{\lambda}(\omega)\equiv1-&\sum^{4}_{i=2}r^{\lambda}_{i-}r^{\lambda}_{i+}e^{-2K_id_i} -\sum^{3}_{i=2}r^{\lambda}_{i-}r^{\lambda}_{(i+1)+}e^{-2K_id_i-2K_{i+1}d_{i+1}} \nonumber \\ +&r^{\lambda}_{2-}r^{\lambda}_{2+}r^{\lambda}_{4-}r^{\lambda}_{4+}e^{-2K_2d_2-2K_{4}d_{4}} -r^{\lambda}_{2-}r^{\lambda}_{4+}e^{-2K_2d_2-2K_{3}d_{3}-2K_{4}d_{4}}=0,~~\lambda=\alpha,\beta,
	\label{surface-modes-condition}
\end{align}
where 
\begin{align}
	r^{\alpha}_{i+}=\frac{K_{i}-K_{i+1}}{K_{i}+K_{i+1}}=-r^{\alpha}_{(i+1)-},~~r^{\beta}_{i+}=
	\frac{\epsilon_{i+1}K_{i}-\epsilon_{i}K_{i+1}}{\epsilon_{i+1}K_{i}+\epsilon_{i}K_{i+1}}=-r^{\beta}_{(i+1)-},~~ K^2_i=k_\Arrowvert^2-\epsilon_{i}(\omega)\frac{\omega^2}{c^2}.
\end{align}
These conditions coincide with the conditions in~\cite{Zhou}, but with different forms. 

The zero-point energy of all the surface modes in the layers may be written as
\begin{equation}
	E=\sum_{n}\frac{1}{2}\hbar\omega^{\alpha}_{n}+\sum_{n}\frac{1}{2}\hbar\omega^{\beta}_{n},
\end{equation}
where $\omega^{\alpha,\beta}_{n}$ denotes the frequencies satisfying the corresponding conditions in Eq.~(\ref{surface-modes-condition}), and the summation includes a two-dimensional integral for the continuous variable ${{\bf k}_\Arrowvert }$. By the residue theorem or, more specifically, the generalized argument theorem, choosing the counterclockwise integration contour $C$ as the whole imaginary axis with its right infinite semicircle in the complex plane of $\omega$, and by the fact that the poles of $G^{\lambda}(\omega)$ in the complex plane, if any, are independent of the thicknesses of the layers and should be of no physical interest, one may formulate the zero-point energy of the system as~\cite{van Kampen,Milonni,Zhou}
\begin{equation}
	E=\frac{\hbar L^2}{4\pi}\frac{1}{2\pi i}\int_{0}^{\infty}dk_\Arrowvert k_\Arrowvert\oint_{C}\omega\left[\frac{G'^{\alpha}(\omega)}{G^{\alpha}(\omega)}+\frac{G'^{\beta}(\omega)}{G^{\beta}(\omega)}\right]d\omega,
\end{equation}
where the overlapping area of the multilayered system is assumed to be $L\times L=L^2$. The integral along the infinite semicircle drops out~\cite{Milonni}, and using integration by parts, we obtain 
\begin{align}
	E(d_2,d_3,d_4)=&\frac{\hbar L^2}{4\pi^2}\int_{0}^{\infty}dk_\Arrowvert k_\Arrowvert\int_{0}^{\infty}d\xi\sum_{\lambda=\alpha,\beta}\ln G^{\lambda}(i\xi;d_2,d_3,d_4),
	\label{total-zero-point-energy1}
\end{align}
where we have used the substitution $\omega=i\xi$ and rearranged the integral range of $\xi$ by recognizing that, according to the analytic properties of the dielectric constant, $\epsilon_i$ and $K_i$, and then $G^{\lambda}$, are even functions of $\xi$. In the following we shall simply rewrite $G^{\lambda}(i\xi)$ as $G^{\lambda}(\xi)$, and 
\begin{equation}
	K^2_i=k_\Arrowvert^2+\epsilon_{i}(i\xi)\frac{\xi^2}{c^2}
\end{equation}
is understood in the latter notation.

To claim Eq.~(\ref{total-zero-point-energy1}) is the total zero-point energy of the multilayered system, we may analyze its immediate predictions as follows.  

\begin{itemize}

\item[1.]$d_{2,3,4}\rightarrow \infty$:

In this limit the zero-point energy of the surface modes should be zero. Using Eqs.~(\ref{surface-modes-condition}) and~(\ref{total-zero-point-energy1}), we have
\begin{equation}
	E(d_2,d_3,d_4)=0.
\end{equation}
The result also manifests that Eq.~(\ref{total-zero-point-energy1}) does not include the distance-independent contribution.

\item[2.]$d_{2,3}\rightarrow \infty$:

In this limit the situation should be reduced to a three-layer problem.
Applying Eqs.~(\ref{surface-modes-condition}) and~(\ref{total-zero-point-energy1}), we find
\begin{align}
	E(d_2,d_3,d_4)&=\frac{\hbar L^2}{4\pi^2}\int_{0}^{\infty}dk_\Arrowvert k_\Arrowvert\int_{0}^{\infty}d\xi\sum_{\lambda=\alpha,\beta}\ln \left(1-r^{\lambda}_{4-}r^{\lambda}_{4+}e^{-2K_4d_4}\right)\nonumber \\
	&\equiv E(d_4),
\end{align}
where $E(d)$ represents the energy of a three-layer system in which the 
distance of the two corresponding dielectric regions is $d$. For $ d_{2,4}\rightarrow \infty$, $E(d_2,d_3,d_4)= E(d_3)$, which reproduces the result of~\cite{Zhou} under this circumstance.

\item[3.]$d_3\rightarrow \infty$:

In this limit the surface modes of the two surfaces of layer $3$ should decouple from each other, and the situation is reduced to two independent three-layer problems. Using Eqs.~(\ref{surface-modes-condition}) and~(\ref{total-zero-point-energy1}), we have
\begin{align}
E(d_2,d_3,d_4)&=\frac{\hbar L^2}{4\pi^2}\int_{0}^{\infty}dk_\Arrowvert k_\Arrowvert\int_{0}^{\infty}d\xi\sum_{\lambda=\alpha,\beta}\left[\ln \left(1-r^{\lambda}_{2-}r^{\lambda}_{2+}e^{-2K_2d_2}\right) +\ln \left(1-r^{\lambda}_{4-}r^{\lambda}_{4+}e^{-2K_4d_4}\right)\right]\nonumber \\
&= E(d_2)+E(d_4).
\label{limitd3}
\end{align}
Note that due to the exponent suppression, in the circumstances that $1)$ the dielectric constants are of the same order; $2)$ $\epsilon_3\gg\epsilon_{2,4}$, $d_3$ may be effectively treated as infinity if 1) $d_3$ is much larger than the other distances; 2) it is not small compared to $d_{2,4}$.

\item[4.] $\epsilon_2=\epsilon_3=\epsilon_4$:

This special circumstance is equivalent to a three-layer problem. By Eqs.~(\ref{surface-modes-condition}) and~(\ref{total-zero-point-energy1}), the energy reads
\begin{align}
E(d_2,d_3,d_4)&=\frac{\hbar L^2}{4\pi^2}\int_{0}^{\infty}dk_\Arrowvert k_\Arrowvert\int_{0}^{\infty}d\xi\sum_{\lambda=\alpha,\beta}\ln(1-r^{\lambda}_{2-}r^{\lambda}_{4+}e^{-2K_2d_2-2K_{3}d_{3}-2K_{4}d_{4}})\nonumber \\
&=E(d_2+d_3+d_4),
\end{align}
where $K_2=K_3=K_4$. 

\end{itemize}

We thus conclude that Eq.~(\ref{total-zero-point-energy1}) represents indeed the total zero-point energy of the five-layer system. In the appendix we shall discuss a way to find the total energy of n-number layers $E(d_2,d_3,...,d_{n-1})$. It should be noted that from the total energy, we can find the normal force between two layers $i-1$ and $i+1$ by
\begin{equation}
F^N_{i-1i+1}=\frac{\partial E(d_2,d_3,...,d_{n-1})}{\partial d_{i}}, 
\end{equation}
where $i=2,3,...,n-1$. One may confirm that the normal force $F^N_{24}=\partial E(d_2,d_3,d_4)/\partial d_3$ is in agreement with the force obtained in~\cite{Zhou}. 

After obtaining the total energy of the multilayered system, we are prepared to calculate the energy difference between the two situations that the appearance and absence of plate C between slabs A and B. One might realize that there are two independent three-layer systems when plate C does not lie between the two slabs, and also that the physical picture requires all the other places are occupied by the same medium ($\epsilon_2=\epsilon_4$). Therefore, the energy difference is

\begin{align}
	E&\equiv E(d_2,d_3,d_4)-E(d_2+d_3+d_4)-E(d_3)\nonumber \\&=\frac{\hbar Lb}{4\pi^2}\int_{0}^{\infty}dk_\Arrowvert k_\Arrowvert\int_{0}^{\infty}d\xi\sum_{\lambda=\alpha,\beta}\left[\ln G^{\lambda}(\xi;d_2,d_3,d_4)-\ln G^{\lambda}(\xi;d_2+d_3+d_4)-\ln G^{\lambda}(\xi;d_3)\right],
	\label{total-zero-point-difference}
\end{align}
where $G^{\lambda}(\xi;d_2+d_3+d_4)=G^{\lambda}(\xi;d_2,d_3,d_4)\arrowvert_{\epsilon_2=\epsilon_3=\epsilon_4}$ and $G^{\lambda}(\xi;d_3)=G^{\lambda}(\xi;d_2,d_3,d_4)\arrowvert_{d_{2,4}\rightarrow \infty}$. 

The difference is in general not zero, and following the discussion given in section II, we obtain the tangential force experienced by plate C in the misaligned system
\begin{align}
	F_T(d_2,d_3,d_4)=-\frac{\partial E}{\partial b}=-\frac{\hbar L}{4\pi^2}\int_{0}^{\infty}dk_\Arrowvert k_\Arrowvert\int_{0}^{\infty}d\xi\sum_{\lambda=\alpha,\beta}\left[\ln G^{\lambda}(\xi;d_2,d_3,d_4)-\ln G^{\lambda}(\xi;d_2+d_3+d_4)-\ln G^{\lambda}(\xi;d_3)\right].
	\label{tangential-force-dielectric1}
\end{align}
Notice that the force is induced by the zero-point energies of the surface modes in the different configurations, and the layer-thickness-independent or non-zero-point-energy contribution, if any, is not considered physical and not calculated (We wish we could).

For convenience, one can also write the force in the form
\begin{align}
	F_T(d_2,d_3,d_4)=-\frac{\hbar L}{4\pi^2c^2}\int_{0}^{\infty}d\xi \xi^2\epsilon_4\int_{1}^{\infty}dpp\sum_{\lambda=\alpha,\beta}\left[\ln G^{\lambda}(\xi;d_2,d_3,d_4)-\ln G^{\lambda}(\xi;d_2+d_3+d_4)-\ln G^{\lambda}(\xi;d_3)\right],
		\label{tangential-force-dielectric2}
\end{align}
where the variable $p$ is introduced by~\cite{Lifshitz}
\begin{align}
 k^{2}_{\Arrowvert}=\epsilon_4\frac{\xi^2}{c^2}(p^2-1),
\end{align}
and then
\begin{align}
	K_i=\sqrt{\epsilon_4}\frac{\xi}{c}s_i,~~s_{i}=\sqrt{\frac{\epsilon_{i}}{\epsilon_4}-1+p^2},~~r^{\alpha}_{i+}=\frac{s_{i}-s_{i+1}}{s_{i}+s_{i+1}},~~r^{\beta}_{i+}=
	\frac{\epsilon_{i+1}s_{i}-\epsilon_{i}s_{i+1}}{\epsilon_{i+1}s_{i}+\epsilon_{i}s_{i+1}},
	\label{new-v}
\end{align}
where $s_4=p$. Though we generally assume the equivalence between a physically-motivated multiple integral and its related iterated integrals, it is still worth noting here that for the improper integrals like the one in Eq.~(\ref{total-zero-point-energy1}) or Eq.~(\ref{tangential-force-dielectric2}), the order of integration is not important in the calculation, since any integrable improper double integral would be absolutely integrable, and then, by Fubini's theorem, in the iterated integrals we are allowed to exchange the order of integration at will.

We next consider some special cases for the tangential force.

Let $d_2\rightarrow \infty$. The tangential force would be
\begin{align}
	F_T(d_3,d_4)&=-\frac{\hbar L}{4\pi^2}\int_{0}^{\infty}dk_\Arrowvert k_\Arrowvert\int_{0}^{\infty}d\xi\sum_{\lambda=\alpha,\beta}\left[\ln G^{\lambda}(\xi;d_3,d_4)-\ln G^{\lambda}(\xi;d_3)\right] \nonumber \\
	&=-\frac{\hbar L}{4\pi^2}\int_{0}^{\infty}dk_\Arrowvert k_\Arrowvert\int_{0}^{\infty}d\xi\sum_{\lambda=\alpha,\beta}\left[\ln\left( 1-\sum^{4}_{i=3}r^{\lambda}_{i-}r^{\lambda}_{i+}e^{-2K_id_i} -r^{\lambda}_{3-}r^{\lambda}_{4+}e^{-\sum^{4}_{i=3}2K_id_i}\right)- \ln\left(1-r^{\lambda}_{3-}r^{\lambda}_{3+}e^{-2K_3d_3}\right)\right],	
	\label{tangential-force-d_2-infty}
\end{align}
where $G^{\lambda}(\xi;d_3,d_4)=G^{\lambda}(\xi;d_2,d_3,d_4)\arrowvert_{d_{2}\rightarrow \infty}$. 

Assume also $d_3$ is a (relatively) large distance and can be treated effectively as infinity [see the remark below Eq.~(\ref{limitd3})]. Then the tangential force in Eq.~(\ref{tangential-force-d_2-infty}), in terms of the variable $p$, is reduced to
\begin{align}
	F_T(d_4)&=-\frac{\hbar L}{4\pi^2c^2}\int_{0}^{\infty}d\xi \xi^2\epsilon_4\int_{1}^{\infty}dpp\sum_{\lambda=\alpha,\beta}\ln G^{\lambda}(\xi;d_4)
	 \nonumber \\
	   &=-\frac{\hbar L}{4\pi^2c^2}\int_{0}^{\infty}d\xi \xi^2\epsilon_4\int_{1}^{\infty}dpp\sum_{\lambda=\alpha,\beta}\ln \left( 1-r^{\lambda}_{4-}r^{\lambda}_{4+}e^{-2\sqrt{\epsilon_4}\xi pd_4/c}\right).
	   \label{two-thick-plate-withd4}
\end{align}
This is the force caused by two parallel but misaligned thick plates with distance $d_4$. 

From now on, we assume $\epsilon_1=\epsilon_3=\epsilon_5\equiv\epsilon$ ($s_1=s_3=s_5\equiv s$) and $\epsilon_2=\epsilon_4=1$ ($s_2=s_4=p$), i.e., the situation that three identical plates are separated by vacuum. We introduce $D^{\lambda}$ by

\begin{align}
D^{\lambda}\equiv r^{\lambda}_{2-}r^{\lambda}_{2+}=r^{\lambda}_{3-}r^{\lambda}_{3+}=r^{\lambda}_{4-}r^{\lambda}_{4+}=r^{\lambda}_{2-}r^{\lambda}_{4+}=-r^{\lambda}_{2-}r^{\lambda}_{3+}=-r^{\lambda}_{3-}r^{\lambda}_{4+},~~\lambda=\alpha,\beta,
\end{align}
where
\begin{align}
	D^{\alpha}=\frac{(s-p)^2}{(s+p)^2},~~D^{\beta}=\frac{(s-\epsilon p)^2}{(s+\epsilon p)^2}.
\end{align}
Then, for example, Eq.~(\ref{two-thick-plate-withd4}) reads
\begin{align}
	F_T(d_4)=-\frac{\hbar L}{4\pi^2c^2}\int_{0}^{\infty}d\xi \xi^2\int_{1}^{\infty}dpp\sum_{\lambda=\alpha,\beta}\ln \left( 1-D^{\lambda}e^{-2\xi pd_4/c}\right).
	\label{D-two-thick-plate-withd4}
\end{align}
In the perfectly conducting limit $\epsilon\rightarrow \infty$ ($D^{\lambda}=1$), 
\begin{align}
	F_T(d_4=d/2)&=-\frac{\hbar c }{2\pi^2d^3}L\int_{0}^{\infty}dxx^2\ln \left( 1-e^{-x}\right)\nonumber \\
	        &=\frac{\pi^2\hbar c }{90d^3}L,
	\label{D=1-two-thick-plate-withd4}
\end{align}
where $x=\xi pd/c$, and we have exchanged the order of integration, i.e., using the substitution $dx=p d/c d\xi$. One may also calculate the integral by the original order, i.e., $dx=\xi d/c dp$ [see also the remark below Eq.~(\ref{new-v})]. As expected, this force coincides with Eq.~(\ref{2}).

\subsection{Finite conductivity}
In the real world, even for good conductors, one may need to consider the finite-conductivity correction to the Casimir effect~\cite{Lifshitz, Hargreaves,Schwinger,Milonni}.  As an application of Eq.~(\ref{tangential-force-dielectric1}), we shall calculate the finite-conductivity corrections for imperfectly conducting plates. 

To simplify the discussion, we consider still that one slab is moved to infinity ($d_2\rightarrow \infty$ say), and assume that the thickness of metal plate $C$ is significantly larger than the effective skin depth approximated by $c/\omega_p$~\cite{Pitaevskii} [$\omega_p=(4\pi Ne^2/m_e)^{\frac{1}{2}}$ is the plasma frequency and $N$ is the number density of free electrons], which means plate C may be considered as an infinitely thick plate ($d_3\rightarrow \infty$). In practice, for a metal plate, $d_3\rightarrow \infty$ would be a satisfied approximation if $d_3\gtrsim 10d_4$~$\sim10~\mu$m.  

For our purpose, the dielectric constant may be approximated as
\begin{equation}
	\epsilon(\omega)=1-\frac{\omega^2_p}{\omega^2}.
	\label{plasma}
\end{equation}
For $D^{\lambda}$ the deviations to the perfectly conducting limit ($D^{\lambda}=1$) are small for the distances $\sim1~\mu$m~\cite{Schwinger}, and the tangential force can be calculated by the perturbation expansion in $\xi/\omega_p$~\cite{Milonni}: 
\begin{align}
	D^{\alpha}=1-4p\frac{\xi}{\omega_p}+8p^2\frac{\xi^2}{\omega^2_p}+\cdots,~~D^{\beta}=1-\frac{4}{p}\frac{\xi}{\omega_p}+\frac{8}{p^2}\frac{\xi^2}{\omega^2_p}+\cdots. 
	\label{order}
\end{align}

Substituting Eq.~(\ref{order}) into Eq.~(\ref{D-two-thick-plate-withd4}), to the first order in $\xi/\omega_p$, one may find the tangential force
\begin{align}
	F_{T}(d_4=d/2)&=\frac{\pi^2\hbar c }{90d^3}L-\frac{\hbar }{\pi^2c^2\omega_p}L\int_{0}^{\infty}d\xi\xi^3\int_{1}^{\infty}dp(1+p^2)\frac{1}{e^{\xi pd/c}-1}\nonumber\\ 
	&=\frac{\pi^2\hbar c }{90d^3}L(1-8\frac{c}{\omega_pd}).
	\label{c-correction}
\end{align}
The second-order correction to the tangential force may be found after simple calculation as
\begin{equation}
	\frac{16\pi^2\hbar c^3}{25\omega^2_pd^5}L.
\end{equation}

\subsection{Finite temperature}

The tangential force discussed above is the force at zero temperature. At finite temperature, the thermal contribution comes into play~\cite{Lifshitz,Sauer,Mehra,Schwinger,Brown}. The significant contribution would come from high temperatures, which is also relevant to the distance between the plates, and for the distances $\lesssim 1~\mu$m, the temperature correction for the Casimir effect might be ignored, see, for example,~\cite{Milonni}.

It is simple to construct the zero-point energy for finite temperature by adding a thermal contribution~\cite{Milonni} 
\begin{equation}
	\text{coth}\frac{\hbar\omega}{2k_BT}=\text{coth}\frac{i\hbar\xi}{2k_BT},
	\label{T}
\end{equation}
where $k_B$ is the Boltzmann constant and $T(\neq0)$ is the temperature (In this place $T$ might not be confused with the subscript $T$ for the tangential forces). Then the energy difference in Eq.~(\ref{total-zero-point-difference}) with the variable $p$ becomes
\begin{equation}
	E(d_2,d_3,d_4;T)=\frac{\hbar Lb}{4\pi^2c^2}\int_{0}^{\infty}d\xi \xi^2\int_{1}^{\infty}dpp\sum_{\lambda=\alpha,\beta}\left[\ln G^{\lambda}(\xi;d_2,d_3,d_4)-\ln G^{\lambda}(\xi;d_2+d_3+d_4)-\ln G^{\lambda}(\xi;d_3)\right]\coth\frac{i\hbar\xi}{2k_BT},
	\label{T-zpd-1}
\end{equation}
where $\epsilon_{2,4}=1$ is assumed.
There are poles in the thermal contribution at 
\begin{equation}
\frac{2n\pi k_BT}{\hbar}=\xi_n,~~n=0,1,2,\cdots,	
\end{equation}
and we need to consider only the contributions from these poles (including the $n=0$ term)~\cite{Lifshitz} due to the observation that the integrand in the analytic region will contribute a purely imaginary number which has no physical meaning. In this treatment $\xi_{n=0}$ may be considered as some sufficiently small quantity. One can perform the integral below the poles with
infinitesimal quarter- or semi-circles in the complex plane, and the energy difference at finite temperature is found as
\begin{align}
	E(d_2,d_3,d_4;T)=\frac{k_BT Lb}{2\pi c^2}\sum_{n=0}^{\infty}{'}\xi^2_{n}\int_{1}^{\infty}dpp\sum_{\lambda=\alpha,\beta}\left[\ln G^{\lambda}(\xi_{n};d_2,d_3,d_4)-\ln G^{\lambda}(\xi_{n};d_2+d_3+d_4)-\ln G^{\lambda}(\xi_{n};d_3)\right],
	\label{T-zpd-2}
\end{align}
where $\Sigma'$ means that there is $1/2$ weight for the $n=0$ term.

We then obtain the tangential force in the misaligned system at finite temperature
\begin{align}
	F_T(d_2,d_3,d_4;T)=-\frac{k_BT L}{2\pi c^2}\sum_{n=0}^{\infty}{'}\xi^2_{n}\int_{1}^{\infty}dpp\sum_{\lambda=\alpha,\beta}\left[\ln G^{\lambda}(\xi_{n};d_2,d_3,d_4)-\ln G^{\lambda}(\xi_{n};d_2+d_3+d_4)-\ln G^{\lambda}(\xi_{n};d_3)\right].
	\label{Temperature}
\end{align}

Now we consider the systems of perfectly conducting plates. For the ideal conductors ($\epsilon\rightarrow\infty$, $D^{\lambda}= 1$), the tangential force reduces to
\begin{align}
	F_T(d_2,d_3,d_4;T)=-\frac{k_BT L}{2\pi c^2}\sum_{n=0}^{\infty}{'}\xi^2_{n}\int_{1}^{\infty}dpp\sum_{\lambda=\alpha,\beta}\left[\ln G^{\lambda}(\xi_{n};d_2)+\ln G^{\lambda}(\xi_{n};d_4)-\ln G^{\lambda}(\xi_{n};d_2+d_3+d_4)\right].
	\label{Temperature}
\end{align}
This formula can be applied to infinitely thin plates, and thus we here may consider similar configurations like Fig.~\ref{Fig2}(\ref{Fig3}) in which $d_3\rightarrow 0$. We have
\begin{align}
	F_T=F_T(d_2=d/2,d_4=d/2;T)=-\frac{ k_BT L}{\pi d^2}\sum_{n=0}^{\infty}{'}\int_{ny}^{\infty}dxx\ln\frac{(1-e^{-x})^2}{1-e^{-2x}}, 
	\label{RdT}
\end{align}
and 
\begin{align}
	F'_T= F_T(d_2\rightarrow\infty,d_4=d/2;T)=-\frac{ k_BT L}{\pi d^2}\sum_{n=0}^{\infty}{'}\int_{ny}^{\infty}dxx\ln(1-e^{-x}),
	\label{RinftyT}
\end{align}
where $x=\xi_npd/c$ and $y=2\pi k_BTd/(\hbar c)$. The forces in general need to be calculated numerically.

However, the high-temperature contribution for perfectly conducting plates can be easily obtained. In the high-temperature limit ($y\gg 1$) which may also be a large-distance limit, it is adequate to consider only the contribution from the $n=0$ term. We find
\begin{align}
	F_T=1.05\frac{k_BT}{\pi d^2}L, 
\end{align}
and
\begin{align}
	F'_T=0.6\frac{k_BT}{\pi d^2}L,
	\label{n=0}
\end{align}
where the Riemann function $\zeta(3)=1.2$ is used. We see that the inequality in Eq.~(\ref{neq}) still holds in this classical limit (notice that $\hbar\rightarrow0$ also implies $y\gg1$).

\begin{figure}[H]
	\centering
	\includegraphics[width=10cm]{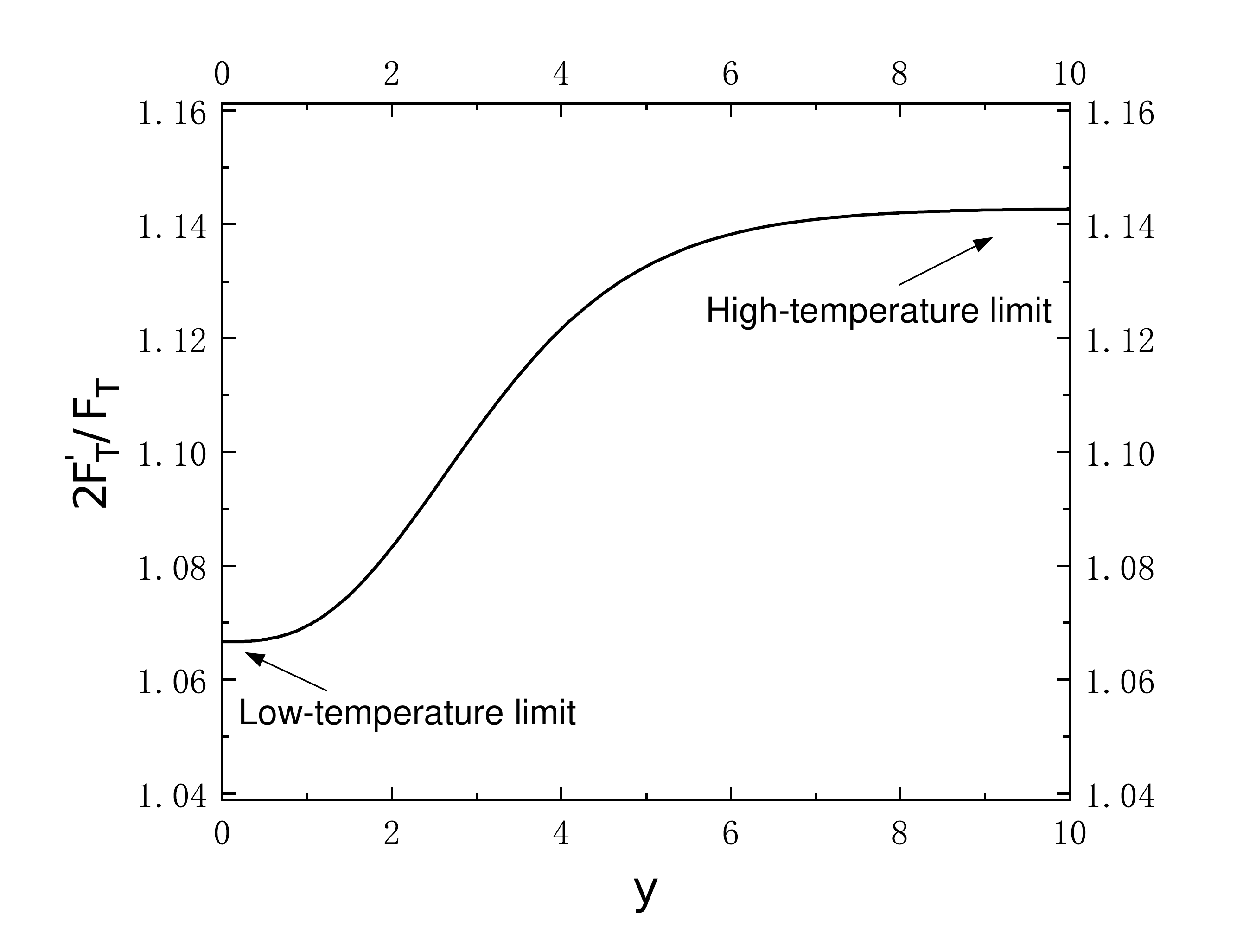}
	\caption{The nonadditivity for the tangential force, characterized by $2F'_T/F_T$, would be reinforced by the finite-temperature corrections.}
	\label{Ratio}
\end{figure}

It is of interest to find approximated analytic formulae for low temperatures. However, the low-temperature limits for the expressions given above are obscure. For such a purpose the Poisson sum method may be used, i.e., the Poisson sum formula~\cite{Mehra, Schwinger} can be used to resum the series in the expressions. We write
\begin{align}
b(n)&\equiv\int_{n y}^{\infty}dxx\ln(1-e^{-x})~\nonumber\\
    &=-\frac{1}{2}n^2y^2\ln{(1-e^{-ny})}-\frac{1}{2}\int_{ n y}^{\infty}dxx^2\frac{1}{e^{x}-1},
\label{bn}
\end{align}
and then we combine the results from~\cite{Mehra, Schwinger} (see also~\cite{Bordag1})

\begin{align}
	\sum_{n=0}^{\infty}{'}b(n)=-\frac{\pi^4}{45}y^{-1}+\frac{1}{720}y^3-\frac{1}{8\pi^2}y^2\sum_{n=1}^{\infty}\frac{1}{n^3}\frac{1+e^{-4\pi^2n/y}}{1-e^{-4\pi^2n/y}}-y\sum_{n=1}^{\infty}\frac{1}{n^2}\frac{e^{-4\pi^2n/y}}{(1-e^{-4\pi^2n/y})^2}.
\end{align}	
For low temperatures ($ y\ll1 $), we have
\begin{align}
	F_T=F_T(T=0)\left\{1+\frac{24\zeta(3)}{\pi^3} \left(\frac{k_BTd}{\hbar c}\right) ^3-\left[\frac{48}{\pi^2}\left(\frac{k_BTd}{\hbar c}\right) ^2+\frac{48}{\pi^3}\left(\frac{k_BTd}{\hbar c}\right)^3\right] e^{-\pi\hbar c /(k_BTd)}         \right\},
\end{align}
and
\begin{align}
	F'_T=F'_T(T=0)\left\{1+\frac{45\zeta(3)}{\pi^3} \left(\frac{k_BTd}{\hbar c}\right)^3-\left(\frac{k_BTd}{\hbar c}\right)^4+\left[\frac{180}{\pi^2}\left(\frac{k_BTd}{\hbar c}\right) ^2 + \frac{90}{\pi^3}\left(\frac{k_BTd}{\hbar c}\right)^3\right]e^{-2\pi\hbar c /(k_BTd)}        \right\},
\end{align}
where the $n=1$ exponentially-suppressed terms are retained. We note here that the finite-temperature corrections to the perfect conductors may reinforce the nonadditivity for the tangential force, see Fig.~\ref{Ratio}.

Before we end this subsection, we may discuss briefly our calculation of Eq.~(\ref{T-zpd-1}). In the standard treatment~\cite{Lifshitz} one considers the contribution from the residue caused by $\xi_{n=0} (\xi=0)$ in the thermal contribution while the integrand seems to have no pole at $\xi=0$, and subsequently we also use the substitutions $x=\xi_npd/c$ in the calculation and $\xi_{n=0}$ is understood as some sufficiently small but non-zero quantity. We, of course, do not concern ourselves with the rigorous mathematics, and however, it might be helpful to examine this situation. We note that the subtlety arising in the situation may be due to the order of integration, since, for the iterated integral in Eq.~(\ref{T-zpd-1}), the different orders of integration might lead to confusions about the actual divergent behaviors of the function of the frequency. To make this point clear, we here calculate Eq.~(\ref{T-zpd-1}) by the different order. To simplify the calculation, we consider the ideal conductor system in which $d_{2,3}\rightarrow\infty$. We then use the prescription that performing the integration over $\xi$ in the last step. Applying the well-defined $x=\xi pd/c$ and $dx=\xi d/cdp$, we have
\begin{align}
	E(d_4=d/2;T)=-\frac{\hbar Lb}{2\pi^2d^2}\int_{0}^{\infty}d\xi\left(\frac{\xi d}{c}\sum_{m=1}^{\infty}\frac{e^{-m\xi d/c}}{m^2}+\sum_{m=1}^{\infty}\frac{e^{-m\xi d/c}}{m^3} \right)\coth\frac{i\hbar\xi}{2k_BT}.
	\label{T-zpd-p2}
\end{align}
Obviously, the integrand now includes a pole at $\xi=0$ in a manifest way, and then the contribution from the $n=0$ ($\xi=0$) emerges naturally in this derivation. However, unfortunately it can be really involved to apply the procedure to the real conductors. Nevertheless, to better understand the thermal behavior of $\xi=0$ with the real metal, the derivation is worthy to be examined for some specific models such as the plasma model and the Drude model~\cite{Bostrom}.

\section{Summary} 
In this work we have discussed the systems that consist of parallel but misaligned finite-size plates from the viewpoint of zero-point energy. We elaborated the zero-point energies of the radiation field in the perfect conductor systems would induce a tangential Casimir force, and the properties and consequences of the tangential force in various conductor systems were studied. For example, an oscillation effect was noticed in an interesting system and a macroscopic nonadditivity for the tangential force was found. For perfect conductors, the precision of the analytic results affected by the geometry was estimated, and it shows that though our approach is an approximation method, the analytic results should be sufficient to provide reliable information on these systems in most accessible geometric regions in experiments. Thereafter, we generalized our study to dielectrics by calculating the total zero-point energies of the surface modes in multilayered dielectrics. The tangential force in the misaligned dielectric systems was found. For imperfectly conducting plates we obtained the finite-conductivity corrections to the tangential force, and we calculated the temperature corrections to the force. We illustrated that the temperature corrections would reinforce the nonadditivity for the tangential force. The strength of the tangential force suggests that it might be observable for moderate experimental settings.

\section*{Acknowledgments}
We would like to thank S. K. Lamoreaux, Tianbo Liu, and P.~W.~Milonni for communications and comments. This project is supported by the National Natural Science Foundation of China (Grant No. 11947032).

\appendix
\section{The total zero-point energy of any-number layers}
In this appendix we outline the way to obtain the total zero-point energy of the surface modes in any-number layers. Starting from Eq.~(\ref{total-zero-point-energy1}), we let $d_4\rightarrow\infty$ which reduces the energy of the system from five layers to four layers, and then let $d_3\rightarrow \infty$ which reduces the energy to the three-layer system. Thereafter, by inductive method, it may be observed that the energy of the surface modes in the $n+1$-layer system can be found from the energy of n-number layers by the following substitution rules: 1) the terms including $r^{\lambda}_{(n-1)+}$: $r^{\lambda}_{(n-1)+}\rightarrow r^{\lambda}_{(n-1)+}+r^{\lambda}_{n+}e^{-2K_nd_n}$; 2) the other terms multiplied by $1-r^{\lambda}_{n-}r^{\lambda}_{n+}e^{-2K_nd_n}$, where the new-added thick layer is located at the $n+1$th place and the original $n$th layer in this circumstance is assumed to be thin \textit{or} the new-added finite-thickness layer is located at the $n$th place and the original $n$th layer which is thick is renumbered as the $n+1$th layer. By this procedure, one may obtain the total energy of any-number layers.


\begin{thebibliography}{99}
\bibitem{Note}	
	Certainly, from the perspective of the van der Waals force, the tangential force cannot be exactly zero as long as the edges of the two plates are still in an asymmetrical position. For example, Ref.~\cite{Wagner} considered the asymmetrical edges in such case with the semitransparent plates and noted that a tangential force can occur due to the asymmetrical edge contributions, see the discussion on the configuration in Fig.~5 in the reference.
\end{thebibliography}
\end{document}